# Nanoscale Photo-Thermal Interaction Theory

Pavel Shafirin[1], Pengli Feng[1], Huanbo Jiang[2], Teri Odom[2], Artur Davoyan[1]

[1]*Mechanical and Aerospace Engineering Department, University of California, Los Angeles, Los Angeles, CA 90095, USA*
[2]*Department of Chemistry, Northwestern University, Evanston, IL 60208, USA*

**ABSTRACT:** Photothermal interaction in nanoscale systems has emerged as a versatile tool for selective heat deposition and robust tuning of optical responses with a myriad of applications from hyperthermal medical treatment to switching and routing of optical signals. However, to date a comprehensive theory of transient photothermal interaction is missing. Development of such a theory is particularly challenged when optical excitation duration is comparable to the timescale related of heat transport – an emergent regime with a mutually interconnected transient interplay of light abortion and heat-induced change of optical responses. Here, we develop a coupled mode theory that captures intricate transient photothermal phenomena in a wide range of nanoscale systems. We further reveal conditions for optimal energy deposition and heating. As an example scenario we apply our model to design metasurfaces with high-contrast and efficient temperature and optical switching, demonstrating fast cooling to the rest state (<40 ns). Beyond exquisite control of both transient temperature profiles and optical responses, our theory offers deep physical insights onto photo-thermal interaction at the nanoscale which can find use in variety of fields from medical treatment to active photonics and manufacturing.

## INTRODUCTION

Optically induced heating at the nanoscale plays an important role in a wide range of applications from nanomedicine [1–3] to photocatalysis [4–8] to energy harvesting [9,10]. Nanostructures not only enhance light—matter interaction [11–13], but also enable highly localized deposition of energy, resulting in steep temperature gradients across extremely short spatial and temporal scales [1,6,14–17]. Importantly, heating and heat diffusion strongly depend on temporal profiles of exciting optical fields [14,15,18,19]. In parallel, thermo-optic nonlinearities alter the refractive indices of the materials involved [20–22], changing both light absorption and the resulting heat generation. As such a complex time-dependent interplay and evolution of both the optical and temperature fields is expected [22–25].

Intertwined coupling between optical fields and transient heating is particularly pronounced when the excitation pulse duration, $\tau_{ex}$, is comparable to that of thermal diffusion to the surrounding medium. The timescale of such "thermalization" to the surroundings is on the order of $\tau_{surr} \sim l^2/D_{surr}$, where $l$ is a characteristic size of the structure and $D_{surr}$ is the thermal diffusivity of surrounding material. As an example, for a silicon nanostructure with $l = 200\ nm$ placed atop a fused silica substrate ($D_{SiO_2} = 8.3 * 10^{-7}\ \frac{m^2}{s}$), $\tau_{surr} \simeq 45\ ns$. Additionally, transient dynamics is amplified in resonant systems, where sharp spectral features become very sensitive to small refractive index variations [20,21,26–28]. To the best of our knowledge, there is no closed-form analytical theory that predicts such transient effects of photo-thermal interaction in nanoscale systems. Nonetheless, photo-thermal phenomena are important and directly affect the performance of various systems, including resonant absorbers [29–31], optical filters [32–34], tunable metasurfaces [21,26,28,35–38], and integrated photonics components [20,39–41], among others.

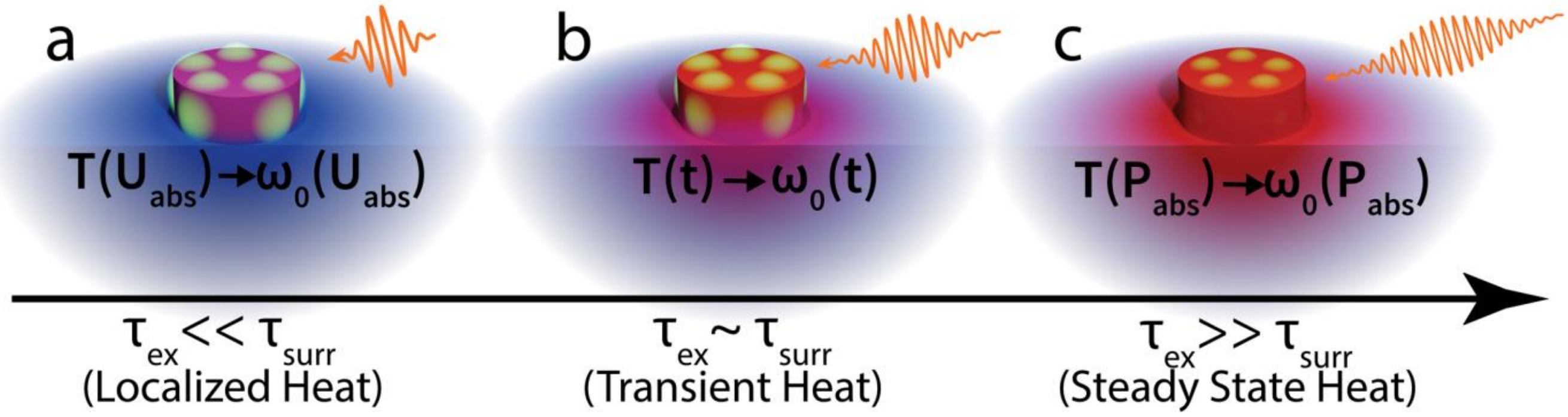

***Figure 1. Transient photo-thermal interaction at the nanoscale.*** ***(a)*** *In a short-pulse excitation regime, when pulse duration $\tau_{ex} \ll \tau_{surr}$, light is absorbed before substantial thermal transport occurs. As such, temperature T and optical mode properties, e.g., resonance frequency $\omega_0$, are functions of absorbed energy $U_{abs}$.* ***(b)*** *When duration of excitation is comparable to a characteristic time of thermal diffusion to surroundings, $\tau_{ex} \simeq \tau_{surr}$, a complex nonequilibrium nonlinear transient interplay is expected.* ***(c)*** *A long-pulse excitation scenario with $\tau_{ex} \gg \tau_{surr}$ results in a quazi-static state evolution of heat, in which temperature and optical properties are slow functions of the incident power.*

Current macroscopic theories of photo-thermal interaction largely focus on two limiting scenarios. The first is a short-excitation regime ($\tau_{ex} \ll \tau_{surr}$, see also Fig. 1a), such as excite by ultrafast laser pulses. In this case light interacts with nanostructures before any substantial thermal diffusion occurs. For sub-picosecond excitation (≤1 ps), the response is dominated by intricate excitation of non-equilibrium carriers that typically thermalize within ~1–10 ps [42–44]. The short lifetime of excited carriers and their related small diffusion length, make optical heating effectively local [45–48], resulting in an initial heat distribution localized near optical absorption hot spots (Fig. 1(a)). In this scenario the resulting thermo-optic interaction is determined by the amount of total absorbed energy $U_{abs}$ [49].

The second limiting scenario is the long-excitation regime ($\tau_{ex} \gg \tau_{surr}$, Fig. 1c), which is governed by a quasi-steady-state temperature distribution established through a balance between thermal diffusion and heat deposition due to light absorption [50]. Such a quasi-steady-state approximation is widely used to model thermo-optic perturbations [51], as well as for predicting self-induced heating [52] and bistability [53,54] in temperature-sensitive nanoresonators under continuous-wave (CW) illumination.

The intermediate regime ($\tau_{ex} \sim \tau_{surr}$, Fig. 1c), however, has received relatively little attention. In this case, neither thermal diffusion can be ignored, nor can a steady-state temperature profile be assumed [22–24]; a fully transient model that couples light-matter interaction with heat transfer is needed. We note that several prior works have examined this intermediate regime assuming optical fields are unaffected by temperature changes [55–57]. In this case optical absorption becomes decoupled from heat transfer and does not capture intricate nonlinear interplay that is anticipated in resonant nanostructures. More recently, finite element models have been employed to study transient photo-thermal effects [22–24]. However, because a fully numerical approach offers little understanding of the underlying mechanisms, it provides limited guidance for designing and optimizing devices of interest. A full self-consistent transient analytical theory of photo-thermal interaction at the nanoscale is yet to be developed.

Here, we derive a coupled-mode theory for transient photothermal interactions in resonant nanophotonic structures. Starting from first principles we capture complex interplay between optical heating and heat driven change of optical properties. We show that optimal heating and light absorption occur at initially detuned resonance frequency. Our generic study is applicable to a diverse range of nanophotonic platforms and accurately captures evolution of heat flow in systems with distinctly different dimensions. Finally, we demonstrate that such a coupled mode theory is well suited for optimizing thermo-optic tuning in metasurfaces. Specifically, we show high contrast and fast control of both metasurface temperature and transmission. Beyond offering a deeper insight into photothermal interaction, our work paves the way for a range of applications, such as optical energy limiters, thermal medical treatment, and pulse forming.

## RESULTS AND DISCUSSION

**Theory of transient photo-thermal interaction.** We begin our study by developing a time-domain coupled-mode theory of photo-thermal interaction in resonant nanoscale structures. Consider a nanostructure resonantly excited by an incident optical pulse of duration $\tau_{ex}$ and field $\boldsymbol{E}_{inc}$. Excitation and subsequent absorption of optical modes inside the nanostructure leads to a temperature rise, which, in turn, stimulates thermo-optic shift of the material's permittivity: $\varepsilon(T) = \varepsilon(T_0) + \delta\varepsilon(T)$ , here $\varepsilon(T_0)$ is the unperturbed permittivity at room temperature, $T_0$, $\delta\varepsilon(T) \simeq 2\sqrt{\varepsilon(T_0)}\alpha\Delta T$ is a temperature induced permittivity shift, $\alpha$ is the thermo-optic coefficient, and $T$ is local temperature [58]. Since, in general, optical excitation and absorption are governed by permittivity, its temperature dependence results in a nonlinear interplay. As such, predicting the resulting temporal evolution of intertwined temperature and optical fields requires solving a mutually coupled system of Maxwell and heat transfer equations. For this purpose, given nanoscale dimensions and slow variation of temperature, we develop an analytical framework based on temporal coupled mode theory (CMT) that predicts photo-thermal responses in a wide range of nanoscale structures.

For the sake of brevity, we provide a detailed first-principles derivation in Supporting Information; here we present only a brief summary of the approach. Assuming a slow variation of optical intensity and temperature fields (as compared to optical oscillation period), we solve Maxwell equations by applying principles of time-domain perturbation theory. In turn, to describe transient temperature evolution we make use of Green's function formalism. After a set of derivations, we obtain the following system of coupled equations (see Supporting Information):

$$\frac{\partial a(t)}{\partial t} + ia(t)\big(\omega_0\big(1-\beta\Delta T_{eff}\big) - i\gamma_0\big) = -\kappa_0 \frac{\partial A_{inc} e^{-i\omega_i t}}{\partial t} \tag{1}$$

$$\Delta T_{eff}(t) = \int_0^t G_{eff}(t,t')|a(t')|^2 dt' \tag{2}$$

Equation (1) describes evolution of a single optical mode resonantly excited within a nanostructure with a field $\boldsymbol{E}_{ex} = a(t)\boldsymbol{E}'_0$, where $a(t) = A(t)e^{-i\omega_0 t - \gamma_0 t}$ is a complex amplitude, $A(t)$ is a slowly varying envelope, $\omega_0$ is the unperturbed resonance frequency, $\gamma_0 = \gamma_R^0 + \gamma_{NR}^0$ is the decay rate, composed of radiative, $\gamma_R^0$, and nonradiative, $\gamma_{NR}^0$, losses, $\boldsymbol{E}'_0$ corresponds to spatial mode profile, $\beta$ is a coefficient connecting the mode's frequency perturbation to temperature: $\beta = \frac{\varepsilon_0}{W}\int_V \alpha\sqrt{\varepsilon(T_0)}|\boldsymbol{E}'_0|^2 dV$, $W = \frac{1}{2}\left[\int_V \varepsilon_0\varepsilon'|\boldsymbol{E}'_0|^2 dV + \int_V \mu_0|\boldsymbol{H}'_0|^2 dV\right]$ with $V$ being the resonator volume that encloses excited optical fields. $A_{inc}$ is the slowly-varying amplitude of the incident field: $\boldsymbol{E}_{inc} = \boldsymbol{E}'_{inc}A_{inc}(t)e^{-i\omega_i t}$, where $\boldsymbol{E}'_{inc}$ is the spatial profile of the incident field and $\omega_i$ is its carrier frequency and $\kappa_0$ is the coupling coefficient. Note that the terms $\omega_0, \gamma_0$ and $\kappa_0$ correspond to the unperturbed, room temperature, mode properties. In Eq. (2) $\Delta T_{eff}$ is an effective temperature increase, which we define as $\Delta T_{eff} = \frac{\varepsilon_0}{\beta W}\int_V \sqrt{\varepsilon'}\alpha\Delta T|\boldsymbol{E}'_0|^2 d\boldsymbol{r}$ and $G_{eff}(t,t')$ is an effective Green's function [59]. Here, for the sake of concept demonstration and without loss of generality, we consider only a single optical mode. The discussion can be extended to a multimode scenario via standard coupled mode formalism [60]. Note that $G_{eff}$ can be calculated with just a single numerical heat transfer simulation by assuming an impulse optical heat source (see Supporting Information for details). As such, Eqs. (1) and (2) provide a simple, yet computationally efficient and powerful tool to predict arbitrary time-domain photo-thermal interaction and evolution inside resonant nanostructures.

For a wide range of nanostructures Eqs. (1) and (2) can be further simplified, so as no numerical simulations are needed beyond standard calculations of the optical mode profile. Specifically, for many nanostructures characteristic time of thermal diffusion across the structure volume, $\tau_V = \frac{l^2}{D_V}$, is rather short as compared to excitation pulse, $\tau_V \ll \tau_{ex}$ (here $D_V$ is the effective thermal diffusivity of the nanostructure). In this case the temperature profile across the resonator volume with a good approximation can be taken to be the solution to a steady state heat equation. If the thermal boundary layer surrounding $V$ is sufficiently large (see Supporting Information for details), the internal temperature and the dynamics of the system is well described by an average temperature increase: $\Delta T_{av} = \frac{1}{V}\int_V \Delta T d\boldsymbol{r}$. The heat transfer equation therefore reduces to (see Supporting Information):

$$\frac{d\Delta T_{av}}{dt} = \frac{1}{V\rho_V c_{p,V}}\left(W|a(t)|^2\gamma_{NR}^0 - \frac{\Delta T_{av} - \Delta T_b}{R_{in}}\right) \tag{3}$$

here $\rho$ and $c_p$ are resonator density and specific heat capacity, $R_{in}$ is a thermal boundary resistance at the surface of a volume $V$ containing an optical mode, $\Delta T_b(t) = \int_0^t \frac{T_{av}-T_b}{R_{in}} G_{surr}(\boldsymbol{r}_0, t|\boldsymbol{r}_0, t')dt'$ is the boundary temperature increase, where $\boldsymbol{r}_0$ sweeps across the volume $V$ (See Supporting information). $G_{surr}$ is the Green's function of the heat equation in the surrounding unbounded medium. For a wide range of nanostructures, owing to their small dimensions and correspondingly quick thermalization across the structure volume, the function $G_{surr}$ can be well approximated by well-known analytical expressions [59].

In turn, the solution to Eq. (1) is drastically simplified when the lifetime of the excited optical mode (i.e., $1/\gamma_0$) is much shorter than the duration of optical pulse and characteristic time of temperature variation ($\left|\frac{dA_{inc}}{dt}\right| \ll \gamma_0|A_{inc}|$ and $\frac{d\Delta T_{av}}{dt} \ll \gamma_0\Delta T_{av}$) – a condition that is typically satisfied for nanoscale structures. In this case quasi-steady-state approximation can be used to find a solution of Eq. (1):

$$a(t) = \frac{\omega_i \kappa_0}{\omega_0 - \omega_i - \beta\Delta T_{eff} - i\gamma_0} A_{inc} e^{-i\omega_i t}. \tag{4}$$

Equations (3) and (4) depend solely on optical mode profile, with all other parameters determined analytically. As such the coupled system of Eqs. (3) and (4) provides a closed-form analytical framework for studying and predicting photo-thermal interaction without a need for complex transient finite element simulations [22–24].

**Photo-thermal interaction in nanophotonic structures.** To demonstrate generic nature of our model, we study three conceptually different nanophotonic structures: a resonant Si metasurface (Fig. 2a), a suspended Si photonic crystal cavity (Fig. 2b), and a Si nanopillar (Fig. 2c) (dimensions of the studied nanostructures can be found in the Supporting Information). We further assume that the metasurface is placed atop $SiO_2$ substrate, the photonic crystal cavity is suspended in air, and the nanopillar is embedded into $SiO_2$ host medium. We consider excitation by a Gaussian laser pulse with a wavelength of $\lambda_{inc} = 1200\,nm$, and amplitude $|A_{inc}|^2 = \frac{U_{pulse}}{\sqrt{2\pi}\tau_{ex}} \exp\left(\frac{-(t-t_0)^2}{2\tau_{ex}^2}\right)$ where $t_0$ is the time at which peak power is reached and $U_{pulse}$ is the pulse energy. We choose $t_0 = 75\,ns$, $\tau_{ex} = 50\,ns$ $U_{pulse} = 0.18, 12.6$ and $0.51\,Jcm^{-1}$ for the cases of metasurface, photonic crystal and nanopillar, respectively.

We first calculate the optical modes using numerical finite difference time domain (FDTD) simulations. The resulting electric field cross-sections for the three structures are shown in the bottom left corners of

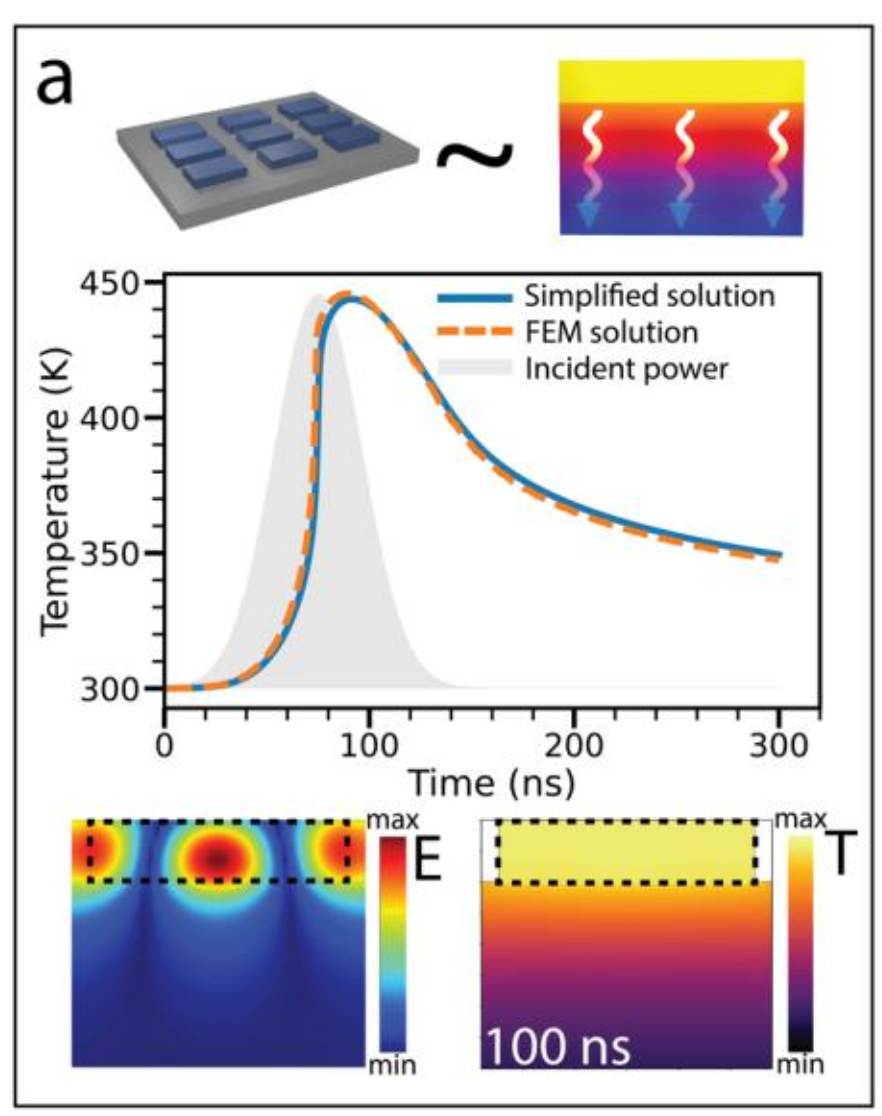

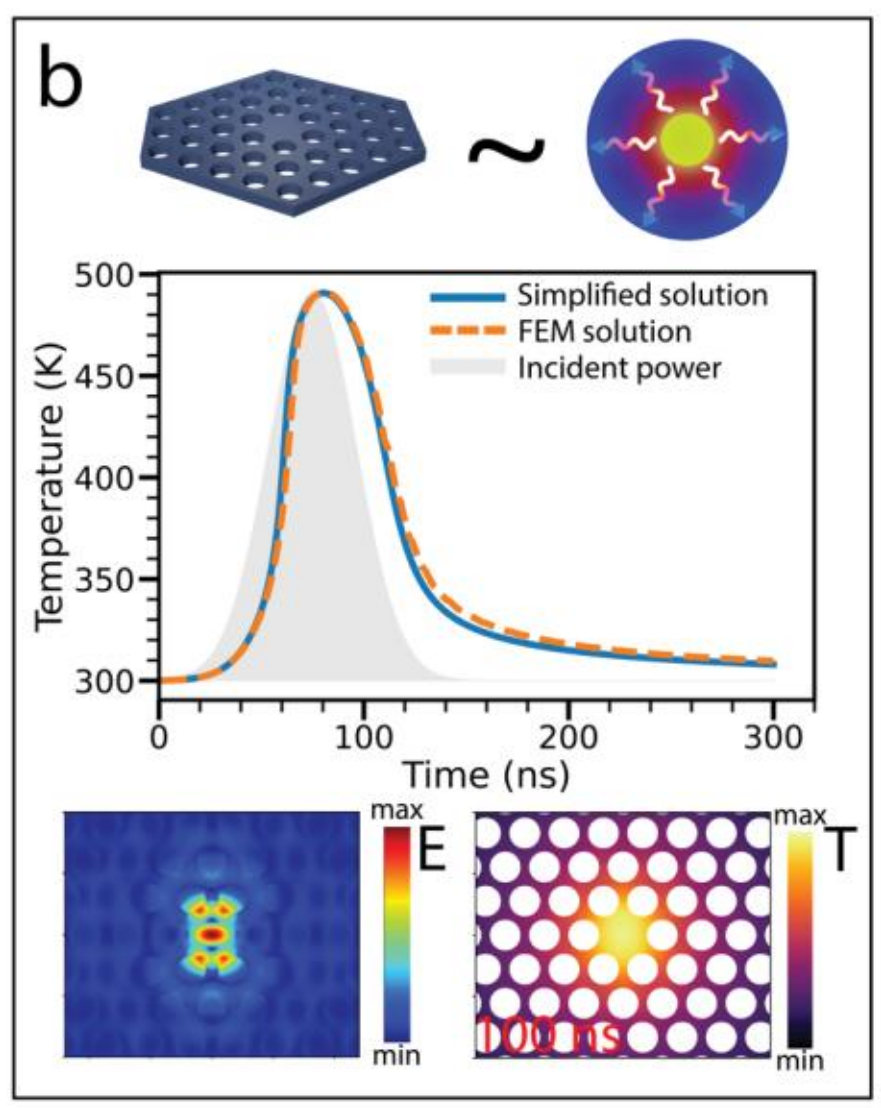

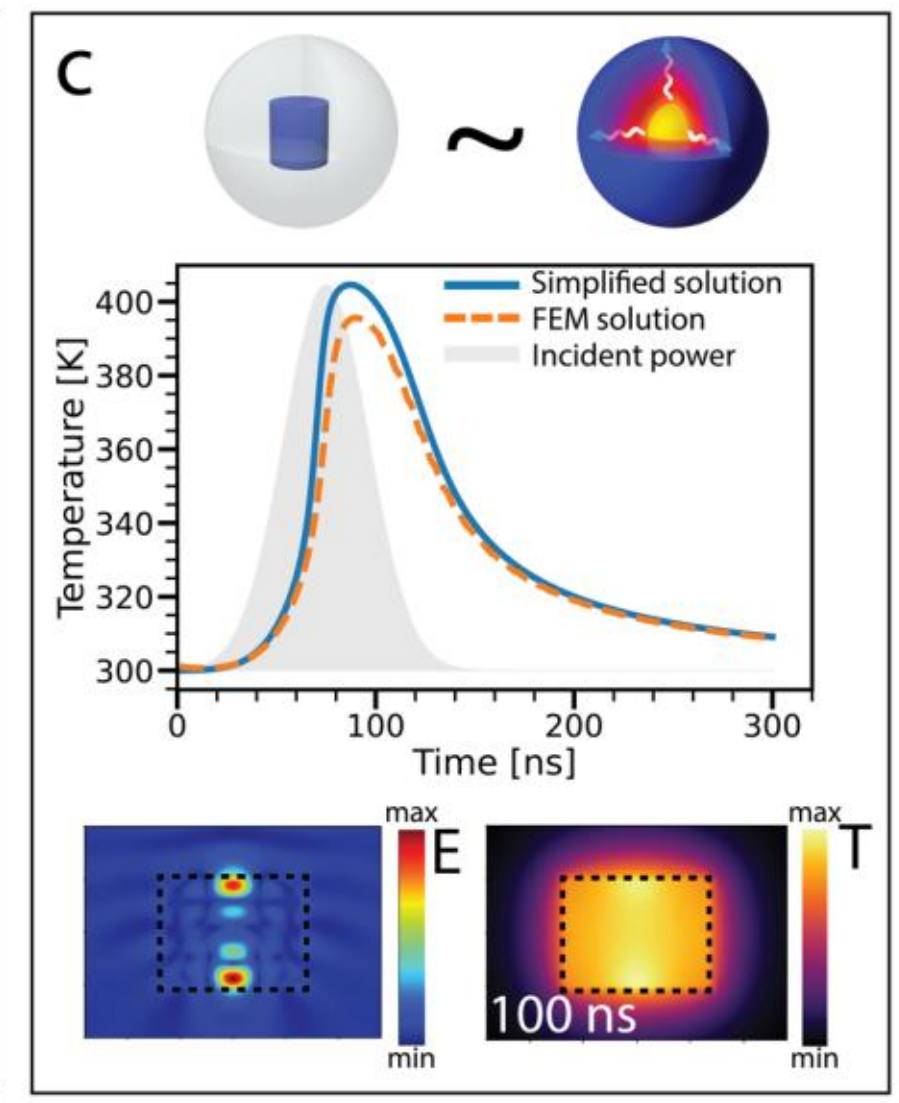

***Figure 2. Transient heating inside resonant nanostructures.*** *(**a**) Temporal evolution of temperature for a Si metasurface atop $SiO_2$ substrate. (**b**) Temporal evolution of temperature inside a resonant cavity formed inside photonic crystal defect. (**c**) Temporal evolution of temperature inside a silicon nanopillar embedded into a silicon dioxide host medium. In all three panels solid blue curve denotes temperature predicted by our analytical model, whereas dashed orange curve corresponds to an average temperatures found by rigorous transient FEM simulation. Optical pulse profile is also shown for comparison (shaded area). Top insets indicted schematics of the geometries studied and their corresponding dimensionality for heat transfer. The bottom left insets show the E-field of the excited mode. Bottom right insets show the temperature distribution at t = 100 ns.*

Figs. 2a, 2b, and 2c, respectively. Each of these example structures can be described by a heat transfer problem of a distinct dimensionality. In particular, for a uniformly illuminated metasurface heat diffusion predominantly occurs along the surface normal direction (i.e., effectively 1D heat transfer problem), for a photonic crystal cavity in-plane heat transport plays a dominant role (i.e., effectively 2D heat transfer problem), whereas for a nanopillar heat is transferred in all 3 directions into the surrounding host medium (i.e., 3D heat transfer problem). Nanoscale size of these structures allows approximating Green's function in Eq. (3) by respective analytical expressions for simple objects [59]: thin film, infinitely long cylinder, and a sphere, respectively (see Supporting Information).

Figures 2a, 2b, and 2c show respective analytical solutions for the evolution of the average resonator temperature $T_{av}$ as a function of time for the three aforementioned structures. The observed dynamics for the studied structures share similarities. Once illuminated, light absorption inside the nanostrutures causes the temperature to rise quickly, reaching a maximum value at a time slightly after the incident laser pulse reaches its peak intensity (i.e., a small lag due to heat diffusion is observed between optical pulse profile and temperature profile). As time passes, heat diffusion causes the temperature to gradually reduce back to initial room temperature value of $300\ K$.

To validate our analytical model, we perform direct spatio-temporal finite element calculations of transient heat equation without any geometry simplifications (similar to [22–24]). The resulting temperature profiles as a function of time are also plotted in Figs. 2a, 2b, and 2c (dashed lines). We also plot cross-sections of the calculated temperature distributions at $t = 100\ ns$ (see right bottom insets in Figs. 2a, 2b, and 2c, respectively). For the metasurface and the photonic crystal, the numerical calculations show that the temperature is nearly uniform across the nanocavity volume, where optical mode is localized. This justifies our assumptions and the use of Eq. (3).

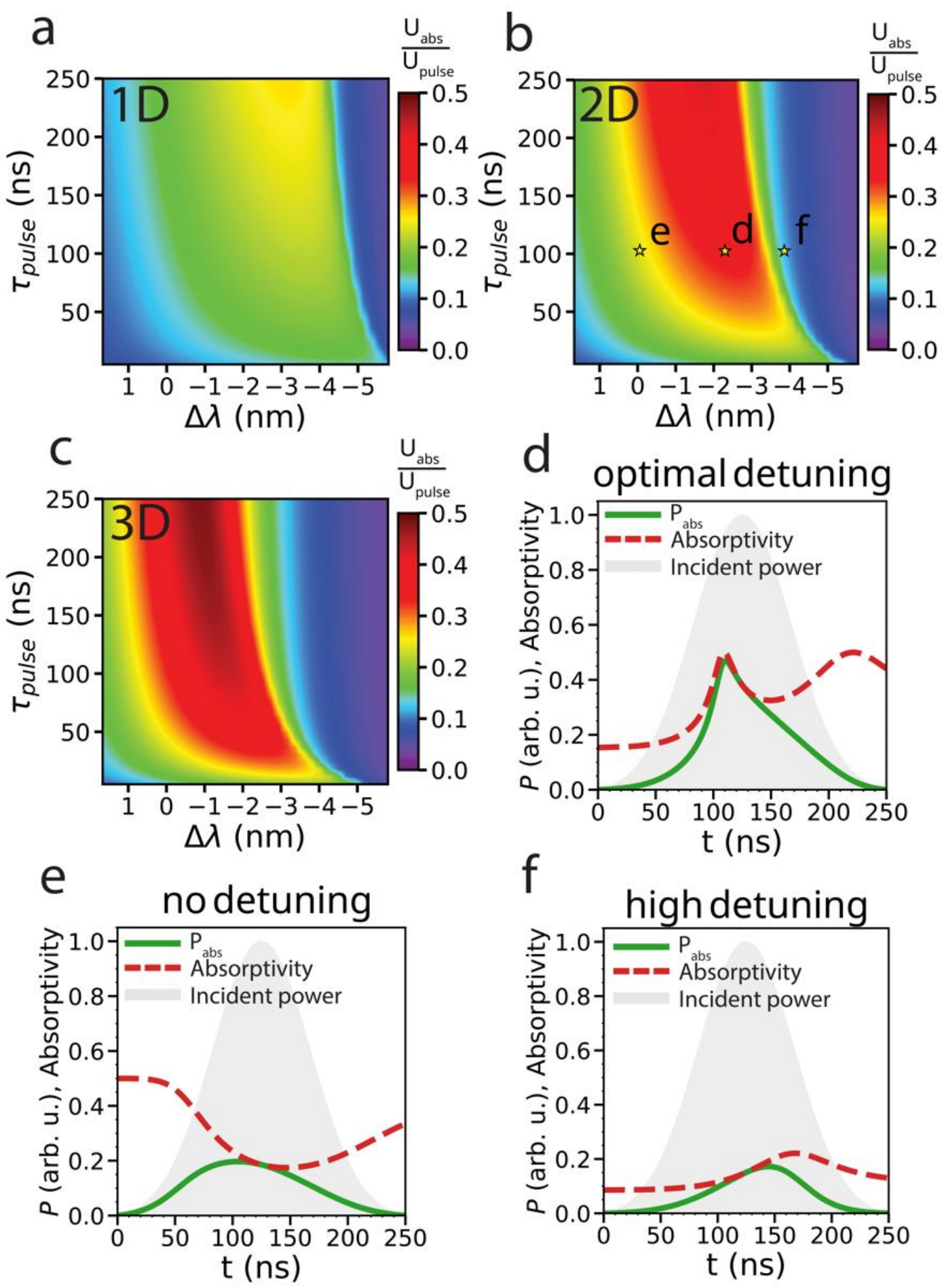

***Figure 3. Parametric studies of absorbed optical power.*** *(**a**) A colormap of the fraction of absorbed laser pulse energy as a function of pulse duration and resonant wavelength detuning calculated for a thin film (1D heat transfer problem). (**b**) A colormap of the fraction of absorbed laser energy as a function of pulse duration and resonant wavelength detuning calculated for a cylinder (2D heat transfer problem). Markers denote regimes studied in panels (d-f). (**c**) A colormap of the fraction of absorbed laser energy as a function of pulse duration and resonant wavelength detuning calculated for a sphere (3D heat transfer problem). (**d**) Transient evolution of absorptivity (dashed red curve) and normalized absorbed power (solid green curve) for zero wavelength detuning and 100 ns pulse duration for a 2D heat transfer problem. (**e**) Same as in panel (d) but for -2.3 nm wavelength detuning. (**f**) Same as in panel (d) but for* $-4\ nm$ *wavelength detuning. In panels (d-f) shaded region indicates the profile of the incident laser pulse.*

We further observe an excellent match between our analytical theory and numerical solutions. A slightly different picture is seen for the case of a nanopillar (Fig. 2c). Owing to its larger size, the electric field distribution of the excited mode across the pillar volume is notably nonuniform, which results in a nonuniform temperature profile across the structure volume. Such nonuniformity leads to a small deviation between our analytical model and numerical solutions. Nonetheless, the analytical solution provides a very good approximation to an exact solution. We note that the solution can be further improved by treating the thermal boundary resistance $R_{in}$ as a fitting parameter to better describe the heat outflow from the nanopillar. Overall, a good agreement between our analytical model and the finite element method (FEM) simulations suggest that complex photo-thermal processes inside a variety of nanophotonic systems are well described and understood by simplified models.

**Efficient energy deposition and heating.** To further elucidate the influence of dimensionality on photo-thermal interaction, we perform a set of parametric studies and explore transient light absorption as a function of resonance detuning and laser pulse duration. We consider a set of simple geometries, such as a thin film, infinite cylinder, and a sphere – which represent 1D, 2D, and 3D space heat transfer problems respectively (these simple structures are well described by respective analytical Green's functions, see Supporting Information). For the sake of concept demonstration, we set the following parameters of the structures fixed: volume $V = 10^{-19} m^3$ , room temperature resonance $\lambda_0 = 1000$ nm, and optical decay coefficients $\gamma_R^0 = \gamma_{NR}^0 = \frac{\pi c}{600\lambda_0}$, where $c$ is the speed of light. In Figs. 3a, 3b, and 3c we plot the fraction of the laser energy absorbed by a nanostructure (i.e., $U_{abs}/U_{pulse}$ ) as a function of pulse duration ($\tau_{pulse}$) and laser detuning from its resonance frequency ($\Delta\lambda = \lambda - \lambda_0$, where $\lambda$ is the laser carrier wavelength). Here, Gaussian pulses with an energy of $U_{pulse} = 4\ nJ$ are assumed.

For all three configurations, we observe a maximum in the total absorbed energy at a negative wavelength detuning (i.e., when $\lambda < \lambda_0$). We attribute this effect to the shift of the structures' resonance with temperature increase (see also derivation of Eq. (1), SI section 1.1). To further explore the dynamics at play, we examine temporal evolution of the instantaneous absorption coefficient for several different laser wavelength detuning values. Taking as reference the cylindrical (2D) case, in Fig. 3d we plot the absorption coefficient and respective normalized absorbed laser power for the case of an optimal detuning (i.e., at which laser absorption inside a nanostructure is maximized). We observe that at initial instances of time the absorption coefficient is low. However, as the structure is heated the resonance shifts, causing the absorption coefficient to peak at around the same time as the incident laser pulse reaches its maximum intensity. A good overlap between the laser peak power and maximum absorption coefficient results in the optimal absorption of incident laser energy by the nanostructure. Conversely, if no laser wavelength detuning ($\Delta\lambda$) is applied (Fig. 3e), the absorption coefficient is initially high but rapidly decreases as the structure heats up and gets detuned from the laser wavelength, leading to suboptimal power absorption. Similarly, if the detuning is too large (Fig. 3f), the structure does not heat sufficiently to compensate for the mismatch, resulting in a consistently low absorbed power.

Further, comparing the fraction of laser power absorbed for the three different dimensions (Figs. 3a, 3b, and 3c, respectively), we observe that as the dimensionality increases the required detuning reduces, while the fraction of power absorbed increases. We attribute this phenomenon to a stronger outgoing heat flux and more efficient cooling as more degrees of freedom are introduced into the heat equation. In addition, for all studied structures we observe that at shorter laser pulse durations a higher initial wavelength detuning is required to reach optimal absorption. Indeed, shorter pulses feature higher peak intensity leading to a much faster heating of the resonator and consequently to higher instantaneous temperatures and resonance shifts.

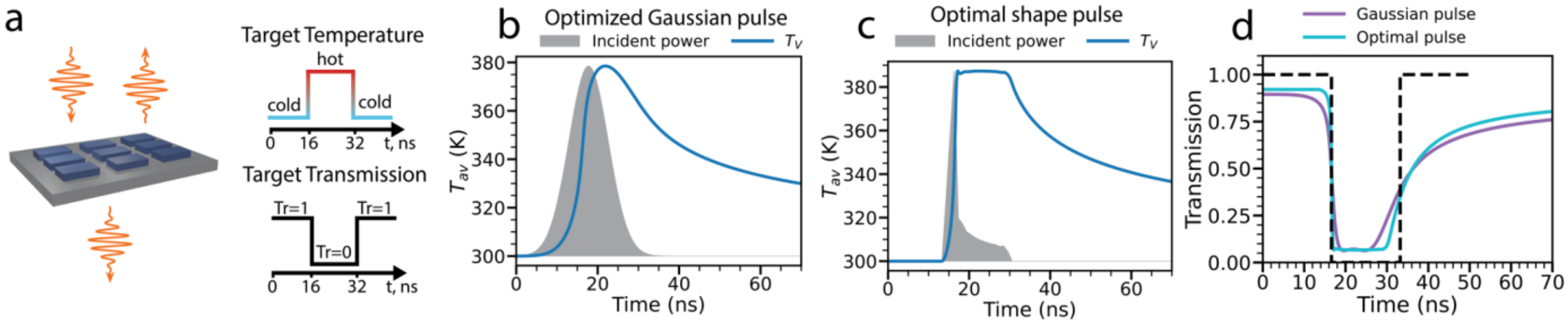

***Figure 4. Ultrafast thermo-optic tuning and temperature control.*** *(**a**) Schematic illustration of a silicon metasurace atop a $SiO_2$ substrate. Insets show idealistic scenario of a desired metasurface switching: when excited metasurface instantaneously reaches desired target operation temperature and stays at that temperature for a desired period of time, as a result, due to the resonance frequency detuning, metasurface transmission switches from near-unity to near-zero. (**b**) Metasurface temperature as a function of time under optimized Gaussian excitation. (**c**) Metasurface temperature as a function of time excited by an optimal-shaped laser pulse. In panels (b) and (c) shaded region indicates the incident beam profile. (**d**) Temporal evolution of metasurface transmission for the two pulses shown in (b) and (c), respectively. Dashed lines indicted target transmission profile.*

**Fast thermal control and switching.** As a final step of our discussion, we explore thermo-optic metasurface tuning [21,22,26,37,38,46]. While thermo-optic tuning offers a robust mechanism for switching optical responses, it is typically limited by a slow response time of thermal diffusion and heat dissipation, which in many traditional systems could reach 10s of microseconds to milliseconds [26,61]. Recently, faster and more efficient tuning has been shown in nanostructures excited by short pulses [22,24,55]. In these systems, shorter pulse duration and small volume of interaction allow for a fast heating of the active volume, while at the same time avoiding parasitic heating of the substrate. A more directed, optimization of transient photo-thermal dynamics can provide additional gains in speed, control precision and energy efficiency of such devices. Additionally, in many applications, such as photo-thermal therapy [1–3] , it may be desirable to achieve a specific constant temperature over a predefined period of time. As we show below, our model is ideally suited for optimizing such transient photo-thermal interactions for controlling both temperature and optical temporal responses.

As an example system, consider again a Si metasurface examined earlier in Fig. 2a. This metasurface structure is of interest as it has low transmission (~7%) when excited at the resonance and is almost entirely transparent when excited far off resonance. As the metasurface resonance depends on temperature, the structure could be easily switched low and high transmission states by heating or cooling. In Fig. 4a we illustrate an ideal scenario of operation. Here, the structure is instantaneously heated from a room-temperature state to a hot state, maintained at this heated level with a constant temperature profile over a predetermined amount of time, and then again instantaneously cooled back to its original room temperature value. In this case, at a specific wavelength the metasurface will switch from a highly transparent one (at room temperature) to mostly opaque (when heated). We aim to find operation conditions that would yield temperature and transmission profiles as close to ideal ones as possible, while using the least amount of laser energy. For this purpose we define a figure of merit as: $FOM = exp\left(\frac{U_{pulse}}{U_{ref}}\right) + \frac{1}{t_{max}}\int_0^{t_{max}}\left(|Tr_{traget}(t) - Tr(t)|\right)dt$. Here the first term creates minimizes cost function for lower pulse energies $U_{pulse}$ with $U_{ref} = 1\, nJ$ being a reference value. The second term measures the difference between the target transmission $Tr_{target}(t)$ shown on Fig. 4(a) and the profile created by our metasurface. Our optimization parameter is the excitation profile $A_i(t)$ as well as the resonator frequency $\omega_0$ (which varies slightly around the excitation frequency). By varying the pulse and resonator frequency we tune the temporal thermal dynamics and switching behavior.

In Figs. 4b and 4d we plot metasurface operation when excited by an optimized Gaussian pulse with a duration of $11.4\, ns$, peak time of $17.8\, ns$ and a fluence of $27.5\, mJ/cm^2$. The optimal resonant frequency is found to be $\omega_0 = 2\pi * 250.44\, rad/s$ which corresponds to an initial detuning of roughly $\Delta\lambda = -3\, nm$ in relation to our probe wavelength of $\lambda_{probe} = 1200\, nm$. Figure 4b shows respective profiles of temperature and laser pulse, whereas Fig. 4d shows a corresponding transmission as a function of time. Initially, the metasurface resonance is detuned from the incident laser wavelength ($\Delta\lambda = 2.3\, nm$), resulting in a 90% transmission. $8\, ns$ after the peak of the pulse, the metasurface temperature peaks at approximately 380 K. Consequently, the resonance wavelength shifts and the metasurface only transmits 7% of the light. As the incident laser power decreases, the metasurface cools down and the transmission gradually goes back to its original state. Evidently, even for an optimal Gaussian pulse maintaining a constant temperature profile in a heated state over a prescribed period of time is not possible. Additionally, use of a Gaussian profile leads to a sub-optimal (i.e., slower) cooling at the trailing edge of the pulse. This is where pulses with optimized temporal profile can be used to further boost performance and efficiency. Our theoretical model is well suited for this purpose.

In Fig. 4c we plot temperature evolution inside a nanostructure for structure illuminated by an pulse with an optimized temporal profile. For a more direct comparison we constrain the energy of this pulse to the same fluence as that of the optimized Gaussian pulse (i.e., $27.5\, mJ/cm^2$). In this case the optimal resonant frequency is found to be $\omega_0 = 2\pi * 250.62\, rad/s$ which corresponds to an initial detuning of roughly $\Delta\lambda = -3\, nm$. The detuning is slightly lager compared to the Gaussian pulse case, which helps achieve a higher switch contrast. Unlike the Gaussian pulse, the optimal-shaped pulse delivers a stronger burst of power on the initial instant of time, which causes rapid heating of the structure to the desired temperature. Laser power is then gradually reduced to balance between light absorption and thermal outflow to surroundings so as to dynamically sustain a near constant temperature (see Eq. (3)). This allows keeping the metasurface heated at $\simeq 385\ K$ over the target $16.6\ ns$ time window. Finally, the pulse shape creates a more favorable condition for faster cooling and return to the original cold state. As a result, the recovery to a transmission of 80% happens within ~40 ns after heating is turned off. We note that here we only optimize the temporal profile of the incident laser pulse, assuming a common metasurface design. A more rigorous co-optimization of the laser shape and of the structure, including its geometry, materials, thermal and optical properties could yield further enhancement of performance. In particular the $SiO_2$ substrate, being a low thermal conductivity material, significantly slows down the thermal recovery. Nonetheless, compared to other thermo-optically tunable devices our model suggests that systems with orders of magnitude improvement in operation speed [22, 26, 61] and switching energies can be designed.

## CONCLUSION

In summary, we have developed a coupled-mode theory of photo-thermal interaction in nanoscale structures. The model provides deep insights into the mechanisms of transient coupling between optical heating and associated heat driven change of optical properties, demonstrating the balance between heat diffusion, light absorption and nanostructure excitation. With the use of our theory structures that maximize laser absorption, maintain near-constant temperature and exhibit fast thermo-optic swithcing can be designed. Our study is applicable to a wide range of systems and will find use in variety of fields from medical treatment to active photonics to manufacturing.

## METHODS

**Semi analytical model solution process.** To solve the semi-analytical model, we develop a Python code that iteratively calculates the resonator temperature from Eqs. (3) and (4). The initial resonator temperature is $T_0 = 300K$, Eq. (3) is then used to produce a temperature at a given time step with a given previous temperature. To obtain the heat flux out of $V$ ($q_{S_V} = (T_V - T_b)/R_{in}$) at each step we apply the Newton-Raphson method to find $T_b$ from Eq. (3) together with the analytical Green's function solution (see SI for details).

**FEM solutions.** The optical simulations are conducted using Lumerical FDTD solutions. The thermal FEM calculations are performed using the Ansys transient thermal solver. In our analytical solutions for the Green's function we assume the following parameters: the metasurface is approximated by a thin film (thickness: $h$ = 118 nm, resonator volume: $V = 3.18 * 10^{-20}\text{m}^3$), the nanopillar is approximated as sphere (radius: $r$ = 667 nm, resonator volume: $V = 1.23 * 10^{-18}\text{m}^3$), the photonic crystal cavity is replaced by a cylinder (radius: $r$ = 226 nm, volume: $V = 4.01 * 10^{-20}\text{m}^3$), here we treat the periodic structure around the cavity as an effective medium to account for lowered heat conduction around the holes. The effective material is calculated by normalizing to the missing volume, which gives the following thermal properties: $\rho_{eff} = 1350.2\ kg^1 m^{-3},\ k_{eff}\ =\ 82.91\ W^1 m^{-1} K^{-1},\ c_p\ =\ 398.8\ J^1 kg^{-1} K^{-1}$.

**Optimization.** The optimization routine is implemented in Python. The optimize.minimize function from the ScyPi library is used to find optimal solutions. When optimizing the pulse shape, we split the relevant time span into 50 points and treat the power at each point as an optimization variable

**Material properties.** The refractive index of silicon at 1200 nm is taken to be $n_{Si}(T\ =\ 300K)\ =\ 3.522$ [62], the thermo-optic coefficient is $\alpha =\ 2\ *\ 10^{-4} K^{-1}$[62], the thermo-optic coefficient of $SiO_2$ is on the order of $10^{-5} K^{-1}$ [63], due to this difference the thermo-optic response of $SiO_2$ is neglected in our analysis. Silicon extinction coefficient is assumed to be $\kappa\ =\ 10^{-3}$, which in practice can be controlled through material doping [64]. Constant thermal properties were assumed for both Si and $SiO_2$: conductivity: $k_{Si} = 148\ W/mK$, $k_{SiO_2} = 10.4\ W/mK$; density: $\rho_{Si} = 2330\ kg/m^3$, $\rho_{SiO_2} = 2650\ kg/m^3$; heat capacity: $c_{p,Si}\ =\ 712\ J/kg$, $c_{p,SiO_2}\ =\ 745\ J/kg$.

ASSOCIATED CONTENT

**Supporting Information.**

The Supporting Information is available free of charge at: …

AUTHOR INFORMATION

Corresponding Author

**Artur Davoyan** - Mechanical and Aerospace Engineering Department, University of California, Los Angeles, Los Angeles, CA 90095, USA; Email: davoyan@ucla.edu

AUTHOR CONTRIBUTIONS

P.S., T.O. and A.R.D initiated the project. P.S. and A.R.D developed the coupled-mode theory. P.S. performed optical and thermal simulations, wrote codes and optimization strategies. P.F. assisted with the thermal models and simulations. All authors contributed to analyzing results and manuscript writing.

**Nanoscale Photo-Thermal Interaction Theory**

**Supporting information**

Pavel Shafirin[1], Pengli Feng[1], Huanbo Jiang[2], Teri Odom[2], Artur Davoyan[1]

[1]*Mechanical and Aerospace Engineering Department, University of California, Los Angeles, Los Angeles, CA 90095, USA*
[2]*Department of Chemistry, Northwestern University, Evanston, IL 60208, USA*

## I. Analytical formalism of photo-thermal interaction

The overall photo-thermal interaction is described by a coupled system of Maxwell's equations and heat diffusion equation, which can be written as follows:

$$\nabla \times \boldsymbol{E} = -\mu_0 \frac{\partial \boldsymbol{H}}{\partial t}, \qquad (1)$$

$$\nabla \times \boldsymbol{H} = \varepsilon_0 \frac{\partial \varepsilon \boldsymbol{E}}{\partial t} + \boldsymbol{J},$$

$$\rho c_p \frac{\partial T}{\partial t} = \nabla(k \nabla T) + Re(\boldsymbol{E}) \varepsilon_0 \frac{\partial Re(\varepsilon)}{\partial t} \qquad (2)$$

where $\varepsilon_0$ and $\mu_0$ are vacuum permittivity and permeability, respectively, $\rho$ is materials density, $c_p$ stands for heat capacity, and $k$ denotes thermal conductivity. In Eq. (1) $\boldsymbol{J}$ is the source current that excites an incident wave with electric and magnetic fields $\boldsymbol{E}_{inc}$ and $\boldsymbol{H}_{inc}$, respectively (e.g., a current at a plane far from the structure of interest that excites an incident plane wave). Fields $\boldsymbol{E}$ and $\boldsymbol{H}$ are total electric and magnetic fields, respectively, which can be represented in the framework of the scattering field formalism as: $\boldsymbol{E} = \boldsymbol{E}_{inc} + \boldsymbol{E}_{ex}$ and $\boldsymbol{H} = \boldsymbol{H}_{inc} + \boldsymbol{H}_{ex}$, where $\boldsymbol{E}_{ex}$ and $\boldsymbol{H}_{ex}$ are electric and magnetic fields excited inside the structure, respectively. Note that here we use a complex field notation. The excited fields, $\boldsymbol{E}_{ex}$ and $\boldsymbol{H}_{ex}$, generally depend on temperature, $T$, through a temperature-dependent dielectric permittivity, $\varepsilon = \varepsilon(T)$. The term $Re(\boldsymbol{E})\varepsilon_0 \frac{\partial Re(\varepsilon \boldsymbol{E})}{\partial t}$ in Eq. (2) is a Joule heating term due to light absorption in the structure [1]. For an optically resonant structure $|\boldsymbol{E}_{ex}| \gg |\boldsymbol{E}_{inc}|$, the heating term can therefore be replaced with $Re(\boldsymbol{E}_{ex})\varepsilon_0 \frac{\partial Re(\varepsilon \boldsymbol{E}_{ex})}{\partial t}$. Both optical fields, $\boldsymbol{E}_{ex}$ and $\boldsymbol{H}_{ex}$, and temperature field, $T$, are functions of space and time. Finding a closed form relation between temperature and excited fields is the main objective of this work.

As mentioned, incident fields $\boldsymbol{E}_{inc}$ and $\boldsymbol{H}_{inc}$ are solutions of the empty space Maxwell equations with a source current $\boldsymbol{J}$:

$$\nabla \times \boldsymbol{E}_{inc} = -\mu_0 \frac{\partial \boldsymbol{H}_{inc}}{\partial t}, \tag{3}$$

$$\nabla \times \boldsymbol{H}_{inc} = \varepsilon_0 \frac{\partial \boldsymbol{E}_{inc}}{\partial t} + \boldsymbol{J},$$

Using Eq. (3) together with scattered field formalism, the Eq. (1) can be written as follows:

$$\nabla \times \boldsymbol{E}_{ex} = -\mu_0 \frac{\partial \boldsymbol{H}_{ex}}{\partial t}, \tag{4}$$

$$\nabla \times \boldsymbol{H}_{ex} = \varepsilon_0 \frac{\partial \varepsilon \boldsymbol{E}_{ex}}{\partial t} + \varepsilon_0 \frac{\partial (\varepsilon - 1)\boldsymbol{E}_{inc}}{\partial t},$$

The permittivity, $\varepsilon$, generally has real and imaginary parts: $\varepsilon = \varepsilon' + i\varepsilon''$, where the imaginary part is responsible for light absorption and heating. For a wide range of materials and problems of interest the imaginary part of the permittivity is much smaller than the real one, i.e., $\varepsilon'' \ll |\varepsilon'|$, which allows treating losses perturbatively. We then present Eq. (4) as follows:

$$\nabla \times \boldsymbol{E}_{ex} = -\mu_0 \frac{\partial \boldsymbol{H}_{ex}}{\partial t}, \tag{5}$$

$$\nabla \times \boldsymbol{H}_{ex} = \varepsilon_0 \frac{\partial \varepsilon' \boldsymbol{E}_{ex}}{\partial t} + i\varepsilon_0 \frac{\partial \varepsilon'' \boldsymbol{E}_{ex}}{\partial t} + \varepsilon_0 \frac{\partial (\varepsilon - 1)\boldsymbol{E}_{inc}}{\partial t},$$

As such photo-thermal interaction problem in photonic nanostructures reduces to solving a coupled system of equations Eq. (2) and Eq. (5).

### I.1. Coupled mode formalism.

Assuming slow variation of temperature as compared to optical oscillation period (i.e., $\frac{dT}{dt} \ll \omega T$, or equivalently $\frac{d\varepsilon}{dt} \ll \omega\varepsilon$, where $\omega$ is a characteristic optical frequency) and assuming that changes in the thermo-optic permittivity are small (i.e., $|\varepsilon(T) - \varepsilon(T_0)| \ll 1$, where $T_0$ is the room temperature), we search for the solutions of Eq. (5) in the framework of the time-dependent perturbation theory. In the scope of this framework the excited fields can be represented as:

$$\begin{pmatrix} \boldsymbol{E}_{ex} \\ \boldsymbol{H}_{ex} \end{pmatrix} = \sum\nolimits_n \begin{pmatrix} \boldsymbol{E}'_n + \delta\boldsymbol{E}'_n \\ \boldsymbol{H}'_n + \delta\boldsymbol{H}'_n \end{pmatrix} A_n(t) e^{-i\widetilde{\omega}_n t} \tag{6}$$

here $\begin{pmatrix} \boldsymbol{E}'_n \\ \boldsymbol{H}'_n \end{pmatrix}$ are time independent spatial electric and magnetic eigenmode field profiles of the lossless unperturbed problem:

$$\nabla \times \boldsymbol{E}'_n = i\widetilde{\omega}_n \mu_0 \boldsymbol{H}'_n, \tag{7}$$

$$\nabla \times \boldsymbol{H}'_n = -i\widetilde{\omega}_n \varepsilon_0 \varepsilon'(T_0) \boldsymbol{E}'_n,$$

index $n$ enumerates eigenmodes ($n = 0, 1, 2, \ldots$), $\widetilde{\omega}_n = \omega_n - i\gamma_R^n$ is a complex eigen-frequency with $\gamma_R^n$ being radiative decay rate of the $n$-th eigenmode (i.e., eigen-frequency is complex for an open problem), $A_n(t)$ are corresponding slowly varying amplitudes (i.e., $\frac{dA_n}{dt} \ll \omega_n A_n$), $\delta\boldsymbol{E}'_n$ and $\delta\boldsymbol{H}'_n$ are small spatial field corrections induced by the photo-thermal perturbations in permittivity, and $\delta\varepsilon(T) = \varepsilon(T) - \varepsilon(T_0)$. Below, for the sake of simplicity and for the purpose of demonstration, we consider a single mode structure. In this case we assume that only $n = 0$ mode is resonantly excited in the nanostructure and write it's complex frequency as $\widetilde{\omega}_n = \omega_0 - i\gamma_R$. We note, however, that the analysis below can be easily extended to a multimode case via standard framework of time-domain perturbation theory [2].

For $n = 0$, system of Eqs. (6) and (7) simplify to:

$$\begin{pmatrix} \boldsymbol{E}_{ex} \\ \boldsymbol{H}_{ex} \end{pmatrix} = \begin{pmatrix} \boldsymbol{E}'_0 + \delta\boldsymbol{E}'_0 \\ \boldsymbol{H}'_0 + \delta\boldsymbol{H}'_0 \end{pmatrix} A_0(t) e^{-i\widetilde{\omega}_0 t} \tag{8}$$

$$-\nabla \times \boldsymbol{E}'_0 = -i\widetilde{\omega}_0 \mu_0 \boldsymbol{H}'_0, \tag{9}$$

$$\nabla \times \boldsymbol{H}'_0 = -i\widetilde{\omega}_0 \varepsilon_0 \varepsilon'(T_0) \boldsymbol{E}'_0,$$

Substituting Eqs. (8) and (9) into Eq. (5) and accounting for slow variations in the first order approximation we obtain the following system of equations:

$$-A_0(t) \nabla \times \delta\boldsymbol{E}'_0 e^{-i\widetilde{\omega}_0 t} = -i\mu_0 \omega_0 \delta\boldsymbol{H}'_0 A_0(t) e^{-i\widetilde{\omega}_0 t} + \mu_0 \frac{\partial A_0(t)}{\partial t} \boldsymbol{H}'_0 e^{-i\widetilde{\omega}_0 t}, \tag{10}$$

$$A_0(t)\nabla\times\delta\boldsymbol{H}_0'e^{-i\widetilde{\omega}_0 t}$$
$$=-i\varepsilon_0\varepsilon'\omega_0\delta\boldsymbol{E}_0'A_0(t)e^{-i\widetilde{\omega}_0 t}-i\varepsilon_0\delta\varepsilon'\omega_0\boldsymbol{E}_0'A_0(t)e^{-i\widetilde{\omega}_0 t}+\varepsilon_0\varepsilon'\boldsymbol{E}_0'\frac{\partial A_0(t)}{\partial t}e^{-i\widetilde{\omega}_0 t}$$
$$+\varepsilon_0\varepsilon''\omega_0\boldsymbol{E}_0'A_0(t)e^{-i\widetilde{\omega}_0 t}+\varepsilon_0\frac{\partial(\varepsilon-1)\boldsymbol{E}_{inc}}{\partial t},$$

In Eq. (10) and subsequent derivations for brevity of notations by $\varepsilon'$ and $\varepsilon''$ we denote $\varepsilon'(T_0)$ and $\varepsilon''(T_0)$, respectively, i.e., one should assume unperturbed permittivity values here.

It is convenient to make a variable substitution $A_0(t)=B_0(t)e^{\gamma_R t}$. In this case the Eq. (10) can be rewritten as:

$$-B_0(t)\nabla\times\delta\boldsymbol{E}_0'e^{-i\omega_0 t}$$
$$=-i\mu_0\omega_0\delta\boldsymbol{H}_0'B_0(t)e^{-i\omega_0 t}+\mu_0\frac{\partial B_0(t)}{\partial t}\boldsymbol{H}_0'e^{-i\omega_0 t}$$
$$+\mu_0\gamma_R B_0(t)\boldsymbol{H}_0'e^{-i\omega_0 t}, \tag{11}$$

$$B_0(t)\nabla\times\delta\boldsymbol{H}_0'e^{-i\omega_0 t}$$
$$=-i\varepsilon_0\varepsilon'\omega_0\delta\boldsymbol{E}_0'B_0(t)e^{-i\omega_0 t}-i\varepsilon_0\delta\varepsilon'\omega_0\boldsymbol{E}_0'B_0(t)e^{-i\omega_0 t}+\varepsilon_0\varepsilon'\boldsymbol{E}_0'\frac{\partial B_0(t)}{\partial t}e^{-i\omega_0 t}$$
$$+\varepsilon_0\varepsilon'\boldsymbol{E}_0'\gamma_R B_0\,e^{-i\omega_0 t}+\varepsilon_0\varepsilon''\omega_0\boldsymbol{E}_0'B_0(t)e^{-i\omega_0 t}+\varepsilon_0\frac{\partial(\varepsilon-1)\boldsymbol{E}_{inc}}{\partial t},$$

Rewriting Eq. (11) as an operator $\mathcal{L}_0=\begin{pmatrix} i\omega_0\varepsilon_0\varepsilon' & \nabla\times \\ -\nabla\times & i\omega_0\mu_0\end{pmatrix}$ acting on the field perturbation $|\delta\boldsymbol{\psi}_0\rangle=\begin{pmatrix}\delta\boldsymbol{E}_0' \\ \delta\boldsymbol{H}_0'\end{pmatrix}$ we obtain:

$$B_0(t)e^{-i\omega_0 t}\mathcal{L}_0|\delta\boldsymbol{\psi}_0\rangle=\bar{\bar{\mathcal{F}}}e^{-i\omega_0 t} \tag{12}$$

where the matrix $\bar{\bar{\mathcal{F}}}$ is given by:

$$\bar{\bar{\mathcal{F}}} = \begin{pmatrix} \begin{matrix} \varepsilon_0\varepsilon' \boldsymbol{E}_0' \dfrac{\partial B_0(t)}{\partial t} + \varepsilon_0\varepsilon' \boldsymbol{E}_0' \gamma_R B_0 + \\ + \varepsilon_0\varepsilon'' \omega_0 \boldsymbol{E}_0' B_0(t) + \varepsilon_0 \dfrac{\partial(\varepsilon - 1)\boldsymbol{E}_{inc}}{\partial t} e^{i\omega_0 t} - \\ - i\varepsilon_0 \delta\varepsilon' \omega_0 \boldsymbol{E}_0' B_0(t) \end{matrix} & 0 \\ 0 & \mu_0 \dfrac{\partial B_0(t)}{\partial t} \boldsymbol{H}_0' + \mu_0 \gamma_R B_0(t) \boldsymbol{H}_0' \end{pmatrix}$$

Per Fredholm alternative theorem, the inhomogeneous equation Eq. (12) has nontrivial solutions only when the adjoint homogeneous problem has a nontrivial solution, i.e., $\mathcal{L}_0^{\dagger}|\boldsymbol{\psi}\rangle = 0$. The adjoint operator to the lossless Maxwell operator $\mathcal{L}_0$ is:

$$\mathcal{L}_0^{\dagger} = \begin{pmatrix} i\omega_0\varepsilon_0\varepsilon' & -\nabla \times \\ \nabla \times & i\omega_0\mu_0 \end{pmatrix}.$$

In the limit of $\gamma_R \ll \omega_0$ the adjoint problem solution is then $|\boldsymbol{\psi}\rangle \simeq \begin{pmatrix} \boldsymbol{E}_0' \\ \boldsymbol{H}_0' \end{pmatrix}^*$, where * denotes complex conjugation. By using definition of Hermitian adjoint operators (i.e., $\langle \boldsymbol{\psi} | \mathcal{L}_0 | \delta\boldsymbol{\psi}_0 \rangle = \langle \delta\boldsymbol{\psi}_0 | \mathcal{L}_0^{\dagger} | \boldsymbol{\psi} \rangle$) from Eq. (11) we obtain:

$$\int_V \begin{pmatrix} \boldsymbol{E}_0' \\ \boldsymbol{H}_0' \end{pmatrix}^* \cdot \bar{\bar{\mathcal{F}}} dV = 0,$$

which can be finally expressed as:

$$\begin{aligned} &\frac{\partial B_0(t)}{\partial t}\left[\int_V \varepsilon_0\varepsilon' |\boldsymbol{E}_0'|^2 dV + \int_V \mu_0 |\boldsymbol{H}_0'|^2 dV\right] - i\omega_0\varepsilon_0 B_0(t) \int_V \delta\varepsilon' |\boldsymbol{E}_0'|^2 dV \\ &\quad + \gamma_R^0 B_0(t)\left[\int_V \varepsilon_0\varepsilon' |\boldsymbol{E}_0'|^2 dV + \int_V \mu_0 |\boldsymbol{H}_0'|^2 dV\right] + \omega_0 B_0(t) \int_V \varepsilon_0\varepsilon'' |\boldsymbol{E}_0'|^2 dV \\ &\quad + e^{i\omega_0 t} \int_V \varepsilon_0 \frac{\partial(\varepsilon - 1)\boldsymbol{E}_{inc}}{\partial t} (\boldsymbol{E}_0')^* dV = 0. \end{aligned}$$

We further denote $W = \frac{1}{2}\left[\int_V \varepsilon_0\varepsilon' |\boldsymbol{E}_0'|^2 dV + \int_V \mu_0 |\boldsymbol{H}_0'|^2 dV\right]$, $\gamma_{NR}^0 = \frac{1}{W}\frac{1}{2}\omega_0\varepsilon_0 \int_V \varepsilon'' |\boldsymbol{E}_0'|^2 dV$, and $\Delta\omega_0 = \frac{1}{2}\frac{\omega_0}{W}\int_V \varepsilon_0 \delta\varepsilon' |\boldsymbol{E}_0'|^2 dV$. With these new notations we obtain:

$$\frac{\partial B_0(t)}{\partial t} - i\Delta\omega_0 B_0(t) + \gamma_R^0 B_0(t) + \gamma_{NR}^0 B_0(t) + e^{i\omega_0 t}\frac{1}{W}\frac{1}{2}\int_V \varepsilon_0 \frac{\partial(\varepsilon-1)\boldsymbol{E}_{inc}}{\partial t}(\boldsymbol{E}'_0)^* \, dV = 0. \tag{13}$$

To further simplify the expression obtained, we represent the incident field in the form of a slowly varying amplitude $\boldsymbol{E}_{inc} = A_{inc}(t)\boldsymbol{E}'_{inc}e^{-i\omega_i t}$, here $\omega_i$ is the carrier frequency, $A_{inc}(t)$ is the slowly varying amplitude of the incident field, and $\boldsymbol{E}'_{inc}$ is the spatial profile of the incident field. In this case Eq. (13) obtains the following form:

$$\frac{\partial B_0(t)}{\partial t} - i\Delta\omega_0 B_0(t) + \gamma_R^0 B_0(t) + \gamma_{NR}^0 B_0(t) + \frac{1}{2}\frac{1}{W}\int_V \varepsilon_0 \frac{\partial(\varepsilon-1)A_{inc}e^{-i(\omega_i-\omega_0)t}}{\partial t}\boldsymbol{E}_{inc}(\boldsymbol{E}'_0)^* \, dV = 0. \tag{14}$$

Next, by writing $B_0(t) = a(t)e^{i\omega_0 t}$, we obtain:

$$\frac{\partial a(t)}{\partial t} + ia(t)(\omega_0 - \Delta\omega_0) + a(t)(\gamma_R^0 + \gamma_{NR}^0) + \frac{1}{2}\frac{1}{W}\int_V \varepsilon_0 \frac{\partial(\varepsilon-1)A_{inc}e^{-i\omega_i t}}{\partial t}\boldsymbol{E}'_{inc}(\boldsymbol{E}'_0)^* \, dV = 0$$

which with the use of slow variation of the fields reduces to:

$$\frac{\partial a(t)}{\partial t} + ia(t)(\omega_0 - \Delta\omega_0) + a(t)(\gamma_R^0 + \gamma_{NR}^0) + \frac{1}{2}\frac{\partial A_{inc}e^{-i\omega_i t}}{\partial t}\frac{1}{W}\int_V \varepsilon_0(\varepsilon-1)\boldsymbol{E}'_{inc}(\boldsymbol{E}'_0)^* \, dV = 0$$

Denoting $\kappa_0 = \frac{1}{2}\frac{1}{W}\int_V \varepsilon_0(\varepsilon-1)\boldsymbol{E}'_{inc}(\boldsymbol{E}'_0)^* \, dV$ we finally obtain:

$$\frac{\partial a(t)}{\partial t} + ia(t)(\omega_0 - \Delta\omega_0 - i\gamma_0) = -\frac{\partial A_{inc}e^{-i\omega_i t}}{\partial t}\kappa_0 \tag{15}$$

where $\gamma_0 = \gamma_{NR}^0 + \gamma_R^0$ is the total loss in the system, $\Delta\omega_0$ corresponds to the shift of the resonance frequency due to photo-thermal interaction, and $\kappa_0$ is a coupling coefficient of the excitation wave to the $0^{th}$ order mode.

The solution to equation (15) can be found using the integrating factor method [3]:

$$a(t) = +\frac{1}{\zeta(t)}\int \zeta(t)\frac{\partial A_{inc}e^{-i\omega_i t}}{\partial t}\kappa_0 dt + C; \quad \zeta(t) = e^{\int i(\omega_0-\Delta\omega_0-i\gamma_0)dt} \tag{16}$$

Assuming that the excitation wave amplitude and the temperature timescales are much larger than the mode lifetime: $\tau_{ex} \gg \gamma_0^{-1}$ and $\left|\frac{dT_{eff}}{dt}\right| \ll \gamma_0|T_{eff}|$. This is typically the case for nanosecond pulses interacting with moderate quality factor resonators ($Q \sim 10^2 - 10^4$) at optical frequencies. The solution reduces to a quasi-static case:

$$a(t) = \frac{\omega_i\kappa_0 A_{inc}}{\omega_0 - \omega_i - \Delta\omega_0 - i\gamma_0}e^{-i\omega_i t} \tag{17}$$

Expression (17) corresponds to optical field evolution in the near-steady-state case. Finally, assuming a constant thermo-optic coefficient $\alpha$, we can express the permittivity perturbation through a temperature increase $\Delta \mathrm{T} = \mathrm{T} - \mathrm{T}_0$ as: $\delta\varepsilon' = 2\sqrt{\varepsilon'(T_0)}\alpha\Delta T$, making the frequency detuning: $\Delta\omega_0 = \frac{\omega_0}{W}\int_V \varepsilon_0\sqrt{\varepsilon'(T_0)}\alpha\Delta T|\boldsymbol{E}_0'|^2 dV$.

I.2. Heat equation

A generic solution of the heat equation, Eq. (2), can be found with the use of Green's function formalism [4]:

$$T(\boldsymbol{r},t) = T_0 + \int_0^t\int_V G(\boldsymbol{r},t,|\boldsymbol{r}',t')\frac{\omega_i}{2}\varepsilon_0\varepsilon''|\boldsymbol{E}'_0(\boldsymbol{r}')|^2|a(t')|^2 d\boldsymbol{r}'dt' \tag{18}$$

where $\frac{\omega_0}{2}\varepsilon_0\varepsilon''|\boldsymbol{E}'_0(\boldsymbol{r}')|^2|a(t')|^2$ is a volumetric heat source due to the light absorption (i.e., $Re(\boldsymbol{E})\varepsilon_0\frac{\partial Re(\varepsilon\boldsymbol{E})}{\partial t} \to \frac{\omega_0}{2}\varepsilon_0\varepsilon''|\boldsymbol{E}'_0(\boldsymbol{r}')|^2|a(t')|^2$ when averaged over an optical period with $\omega_0 \approx \omega_i$), and $G(\boldsymbol{r},t,|\boldsymbol{r}',t')$ is the Green's function of the corresponding heat diffusion problem.

It is convenient to introduce an effective temperature $\Delta T_{eff} = \frac{1}{C_E}\int_V \sqrt{\varepsilon'}\alpha\Delta T|\boldsymbol{E}'_0|^2 d\boldsymbol{r}$, which plays a role of a weighted average temperature across the structure and allows rewriting the frequency shift as $\Delta\omega_0 = \omega_0\beta\Delta T_{eff}$; here $C_E = \int_V \sqrt{\varepsilon'}\alpha|\boldsymbol{E}'_0|^2 d\boldsymbol{r}$ and $\beta = \frac{\varepsilon_0}{W}C_E$. By integrating the solution for temperature given in Eq. (18), we can represent effective temperature as:

$$\Delta T_{eff} = \frac{1}{C_E}\int_V \sqrt{\varepsilon'}\alpha \left[\int_0^t \int_V G(\boldsymbol{r},t,|\boldsymbol{r}',t')\frac{\omega_0}{2}\varepsilon_0\varepsilon''|\boldsymbol{E_0}|^2(\boldsymbol{r}',t')d\boldsymbol{r}'dt'\right]|\boldsymbol{E}'_0|^2 d\boldsymbol{r}$$
$$= \frac{1}{C_E}\int_V \sqrt{\varepsilon'}\alpha|\boldsymbol{E}'_0(r)|^2 \int_0^t \int_V G(\boldsymbol{r},t,|\boldsymbol{r}',t')\frac{\omega_0}{2}\varepsilon_0\varepsilon''|\boldsymbol{E}'_0(\boldsymbol{r}')|^2|a(t')|^2 d\boldsymbol{r}'dt'\,d\boldsymbol{r}$$
$$= \int_0^t |a(t')|^2\left[\frac{1}{C_E}\int_V \sqrt{\varepsilon'}\alpha|\boldsymbol{E}'_0(r)|^2 \int_V G(\boldsymbol{r},t,|\boldsymbol{r}',t')\frac{\omega_0}{2}\varepsilon_0\varepsilon''|\boldsymbol{E}'_0(\boldsymbol{r}')|^2 d\boldsymbol{r}'\,d\boldsymbol{r}\right]dt'$$

The expression in the square brackets can be viewed as an effective Green's function $G_{eff}(t|t')$ that determines temporal evolution of effective temperature, $\Delta T_{eff}$, driven by the optical mode amplitude $a$:

$$\Delta T_{eff}(t) = \int_0^t G_{eff}(t|t')|a(t')|^2 dt' \qquad (19)$$

In complex geometries, Green's function, $G$, does not have a closed form easy-to-use analytical expression, and needs to be computed numerically. At the same time, function $G_{eff}$ can be obtained from a single numerical simulation. In particular, $G_{eff}(t|t')$ is the thermal response to a delta function excitation at time $t'$ with $|a|^2 = \delta(t-t')$. Therefore, numerically simulating an arbitrary geometry with a volumetric source term given by $q''' = \delta(t-t')\frac{\omega_0}{2}\varepsilon_0\varepsilon''|\boldsymbol{E}_0'(\boldsymbol{r}')|^2$, where $\boldsymbol{E}_0'(\boldsymbol{r}')$ is a known eigen-mode profile, and calculating $\Delta T_{eff}$ from the resulting temperature filed yields $G_{eff}$. Such an approach drastically simplifies computations, as any arbitrarily shaped excitation can then be solved for using the Eq. (**Error! Reference source not found.**9).

With the help of $\Delta T_{eff}(t)$ the frequency shift $\Delta\omega_0$ can be expressed as:

$$\Delta\omega_0 = \omega_0\beta\Delta T_{eff} \qquad (20)$$

where $\beta = \frac{\varepsilon_0}{W}C_E$. As such, our final system of equations is:

$$\frac{\partial a(t)}{\partial t} + ia(t)\big(\omega_0(1-\beta\Delta T_{eff}) - i\gamma_0\big) = -\kappa_0\frac{\partial A_{inc}e^{-i\omega_i t}}{\partial t} \qquad (21)$$

$$\Delta T_{eff}(t) = T_0 + \frac{1}{C_E}\int_0^t G_{eff}(t|t')|a(t')|^2 dt' \qquad (22)$$

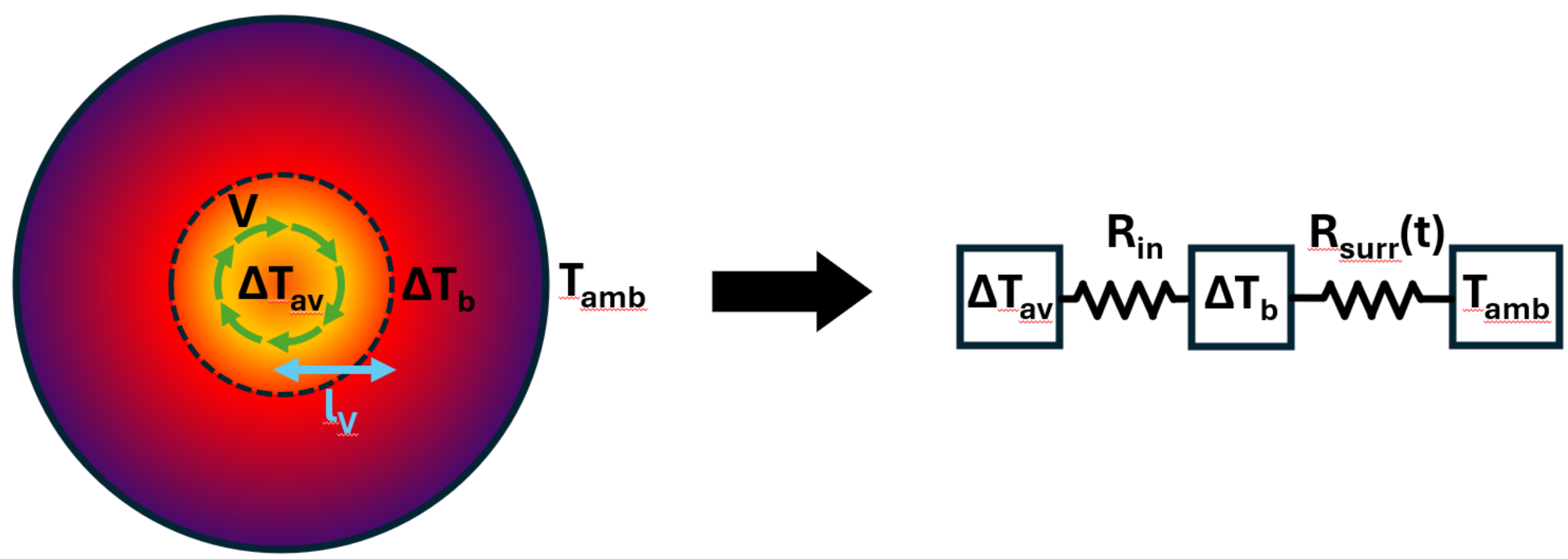

***Figure S1.*** *(Left) Schematic illustration of the thermal profile induced by the optical field. Dashed lines indicate two domains, one with volumetric heating (internal) and one without (external). (Right) A thermal resistance analog of the system at a short time scale. Here the internal heat is close to a steady state, whereas the external heat is described as an expanding boundary layer.*

For many practical calculations, with the use of additional simplifying assumptions the solutions can be searched for in the framework of a lumped model [5]. Consider a volume $V$ (Fig. S1, left) that contains both the excited optical mode and the associated heat source. A characteristic time for the temperature inside $V$ to reach a near steady-state distribution is:

$$\tau_V = \frac{l_V^2}{D_V}$$

where $l_V$ is the characteristic size of $V$, and $D_V = \frac{k_V}{\rho_V c_{p,V}}$ is the effective thermal diffusivity inside the volume $V$. For many nanostructures with high thermal diffusivity, this time is very short (e.g., $\tau_V \approx$ 0.5 ns for a 200 nm silicon structure).

If the excitation pulse duration satisfies $\tau_{\text{ex}} \gg \tau_V$, the temperature inside $V$ is approximately quasi-static. In this case, heat flow can be described using a thermal resistance model. We express the heat flux leaving the volume through an average temperature increase $\Delta T_{\text{av}}$:

$$q_{\text{out}} = \frac{\Delta T_{\text{av}} - \Delta T_b}{R_{\text{in}}}; \qquad \Delta T_{\text{av}}(t) = \frac{1}{V}\int_V \Delta T(\mathbf{r}, t)\, dV, \tag{23}$$

where $\Delta T_b$ is the increase of the boundary temperature above the ambient $T_0$, and $R_{\text{in}}$ is the internal thermal resistance [5] between the bulk and the boundary. While $R_{\text{in}}$ depends on geometry and heat source distribution, its scale can be estimated assuming uniform 1D heating, resulting in $R_{\text{in}} = \frac{l_V}{6 k_V \sigma_V}$, where $\sigma_V$ is the surface area of $V$. We note that variations in source shape introduce only order-unity corrections. For example, if the heat source is an infinitely small point in the center of V, $R_{\text{in}}^{(\text{point})} \approx 1.5 R_{\text{in}}$.

Outside the volume, heat transport is transient and cannot be treated as steady state. However, for times $t \ll \tau_{\mathrm{ex}}$, the temperature field can be approximated as an expanding thermal boundary layer with thickness $\delta_{\mathrm{surr}} = \sqrt{\pi D_{\mathrm{surr}} t}$, where $D_{\mathrm{surr}}$ is the diffusivity of the surrounding material. The corresponding external thermal resistance is $R_{\mathrm{surr}} = \frac{\delta_{\mathrm{surr}}}{k_{\mathrm{surr}} \sigma_V}$. The system can therefore be modeled as two resistances in series (Fig. S1, right). A key simplification occurs when $R_{\mathrm{surr}} \gg R_{\mathrm{in}}$, so that most of the temperature drop occurs in the surrounding medium, and the internal temperature is nearly uniform ($\Delta T_{\mathrm{av}} \approx \Delta T_b$).

This condition is reached for excitation times $\tau_{\mathrm{ex}} \gg \frac{k_{\mathrm{surr}}^2 l_V^2}{36 k_V^2 \pi D_{\mathrm{surr}}}$. It is easily satisfied for highly conductive structures embedded in low-conductivity environments and is similar to the diffusion timescale for materials of the same conductivity. For example, a 200 nm silicon resonator on a $SiO_2$ substrate yields a characteristic timescale below 1 ps.

Under these conditions, the effective temperature can be replaced with the average temperature $\Delta T_{\mathrm{eff}} = \Delta T_{\mathrm{av}}$, and it's dynamics can be obtained by integrating the heat equation over $V$:

$$\frac{d\Delta T_{\mathrm{av}}}{dt} = \frac{1}{V \rho_V c_{p,V}} \left( \int_{\sigma_V} (k \nabla T \cdot \mathbf{n}_V)\, d\sigma_V + \int_V \frac{\omega_0}{2} \varepsilon_0 \varepsilon'' \mid \mathbf{E}_0' \mid^2 \, dV \right). \tag{24}$$

Here, the first term represents heat flux through the boundary and can be approximated as $(\Delta T_{\mathrm{av}} - \Delta T_b)/R_{\mathrm{in}}$. The second term corresponds to optical absorption, which can be written as $W|a(t)|^2 \gamma_{NR}^0$, where $\gamma_{NR}^0 = \frac{1}{2W} \omega_0 \varepsilon_0 \int_V \varepsilon'' |\boldsymbol{E}_0'|^2 dV$.

The resulting temperature evolution equation becomes:

$$\frac{dT_{\mathrm{av}}}{dt} = \frac{1}{V \rho_V c_{p,V}} \left( W|a(t)|^2 \gamma_{\mathrm{NR}}^0 - \frac{T_{\mathrm{av}} - T_b}{R_{\mathrm{in}}} \right). \tag{25}$$

To close the system, we determine the boundary temperature using a Green's function formulation [4]. The surrounding temperature field is:

$$T_{\mathrm{surr}} = T_0 + \int_0^t \int_{\sigma_V} q''(\mathbf{r}_{\sigma_V}, t')\, G_{\mathrm{out}}(\mathbf{r}, t \mid \mathbf{r}_{\sigma_V}, t')\, d\sigma_V dt'. \tag{26}$$

Where $q''$ is the heat flux through the surface of $V$ and $G_{\mathrm{out}}$ is the Green's function of the heat equation for the surrounding volume. Unlike Eq. (18), Eq. (26) involves only the surface heat flux $q''$. If we can then choose a symmetric $V$ such that the flux is uniform along the boundary, the surface integral over $q''$ in Eq. (26) becomes trivial and is equal to the total heat flux leaving $V$, which we have already estimated in Eq. (23). Evaluating Eq. (26) at the boundary ($\boldsymbol{r} = \boldsymbol{r}_{\sigma_V}$) then yields an expression for the boundary temperature:

$$T_b = T_0 + \int_0^t \frac{T_{\mathrm{av}} - T_b}{R_{\mathrm{in}}} G_{\mathrm{out}}(\mathbf{r}_{\sigma_V}, t \mid \mathbf{r}_{\sigma_V}, t')\, dt'. \qquad (27)$$

We note that since under these approximations the majority of the thermal resistance is in the outside region $V$, the specific choice of $R_{in}$ does not meaningfully impact the solution. Therefore, even a very rough estimate of $R_{in}$ is sufficient to produce reliable results.

In this work the temperature dynamics is calculated using Eqs. (17), (25) and (27). The systems we analyze in this work can be effectively reduced to three geometries with known analytical Green's functions [4]:

1D (a planar surface):

$$G_{1D}(0, t \mid 0, t') = \frac{2}{\sqrt{4\pi D(t - t')}}$$

2D (a cylinder of radius $r_0$):

$$G_{2D}(r_0, t \mid r_0, t') = \frac{2}{\pi^2 r_0^2} \int_0^\infty \frac{e^{-\xi D(t-t')/r_0^2}}{\xi[J_1^2(\xi) + Y_1^2(\xi)]}\, d\xi$$

3D (a sphere of radius $r_0$):

$$G_{3D}(r_0, t \mid r_0, t') = \frac{1}{2\pi r_0^2}[4\pi D(t - t')]^{-1/2} - \frac{1}{4\pi r_0^3} e^{-Dt'/r_0^2}\, \mathrm{erf}\left(\frac{\sqrt{D(t - t')}}{r_0}\right)$$

These correspond to planar, cylindrical, and spherical geometries, respectively. As shown in Fig. 3 of the main text, this simplified framework accurately captures temperature dynamics across a wide range of nanophotonic systems on nanosecond timescales.

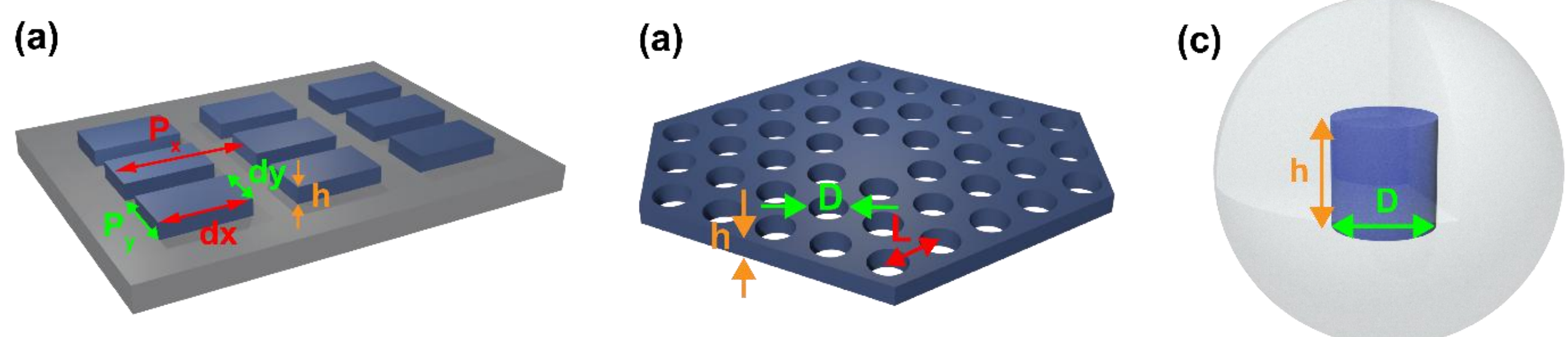

***Figure S5.*** *Illustrations of the modelled optical resonators with dimensions. (a) The periodic metasurface. (b) Photonic crystal cavity. (c) Individual nanopillar scatterer.*

**II. Test Resonant Structures**

We apply our model to three conceptually different resonant nanophotonic structures. The first structure (Fig. S2 a) is a metasurface consisting of periodically arranged Si rectangles on top of a $SiO_2$ substrate. The periods are $P_x = 563\,nm$ and $P_y = 480\,nm$, the rectangle dimensions are $dx = 501\,nm$, $dy = 320\,nm$, $h = 200\,nm$. The metasurface has a resonance at $\lambda_{res}(T = 300K) = 1196nm$ with a quality factor of $Q = 500$.

The second structure (Fig. S2 b) is a suspended Si photonic crystal. The hexagonal array period is $L = 345\,nm$, the hole diameter $D = 240\,nm$ and the height $h = 250\,nm$. The resonance occurs at $\lambda_{res}(T = 300K) = 1192nm$ and has a quality factor of $Q = 157$.

The last structure is a Silicon nanopillar embedded into $SiO_2$. The diameter is $D = 1266\,nm$ and the height $h = 980\,nm$. The room temperature resonance frequency is $\lambda_{res}(T = 300K) = 1195nm$, and the quality factor is $Q = 233$.

The fast switching optimization discussed in the main text is performed for the 1D metasurface system (Fig. S2(a)). Since the performance is strongly affected by the initial resonator frequency $\omega_0$ we treat is as another optimization parameter alongside the excitation pulse shape $A_{inc}$. In practice the resonant frequency can be tuned by slightly adjusting the period of the metasurface array. We use coupled mode theory to calculate the transmission and absorption of the probe beam at any point in the optimization with calculated loss values of $\gamma_R = 1.16$ THz and $\gamma_{NR} = 0.39\,THz$. Figure S3 shows the numerically calculated transmission, reflection and absorption spectra of the metasurface together with the corresponding coupled mode theory values. We consider a probe wavelength of $\lambda_{probe} = 1200\,nm$, consequently at room temperature the metasurface reflects over 90% on incident light. When heated the resonant peak shifts to longer wavelengths, decreasing transmission.

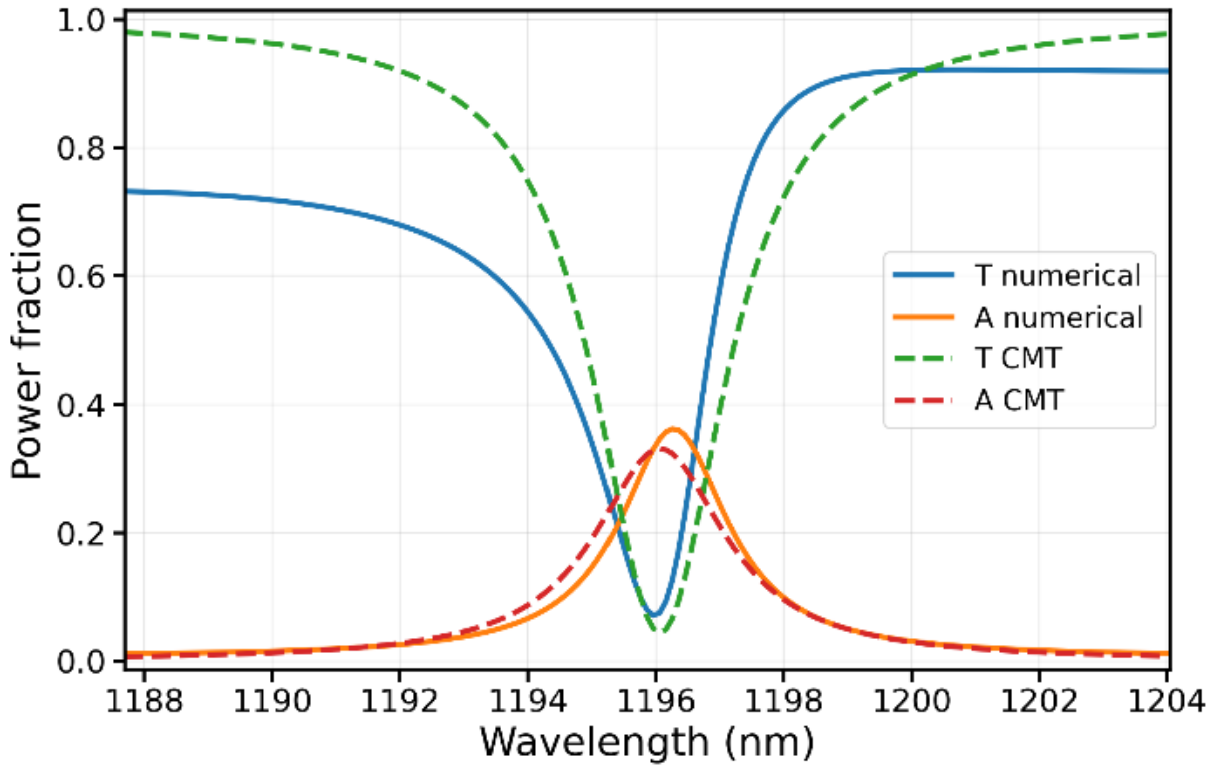

*Figure S3. Numerically calculated transmission and absorption spectra of the metasurface (solid) and the CMT approximation used (dashed).*